\documentstyle[amssymb,aps]{revtex}

\def\XTheorem{1}
\def\XDnondeg{2}
\def\XLnondeg{3}
\def\Xomega{4}
\def\Xdecompo{5}
\def\XVarLemma{6}
\def\XUniqueProp{7}
\def\XStSprThm{8}

\def\Label#1{\label{#1}}
\def\idty{{\leavevmode{\rm 1\ifmmode\mkern -5.4mu\else\kern -.3em\fi I}}}
\def\Ibb #1{ {\rm I\ifmmode\mkern -3.6mu\else\kern -.2em\fi#1}}
\def\ibb #1{\leavevmode\hbox{\kern.3em\vrule
     height 1.5ex depth -.1ex width .2pt\kern-.3em\rm#1}}
 \def\Cx {{\ibb C}} 
\def\Bar{\overline}
\def\tr {\mathop{\rm tr}\nolimits}
\def\rstr{\hbox{$\vert\mkern-4.8mu\hbox{\rm\`{}}\mkern-3mu$}} 
\def\phi{\varphi} 
\def\epsilon{\varepsilon} 
\def\3{\ss} 
\def\B{{\cal B}} 
\def\H{{\cal H}} 
\def\K{{\cal K}} 
\def\SU#1{{\rm SU}(#1)} 
\def\Lie#1{{\frak{\lowercase{#1}}}}

\renewcommand{\Bbb}[1]{\if1#1\idty\else{\mathbb #1}\fi}
\newcommand{\goth}[1]{{\frak #1}}
\newcommand{\operatorname}[1]{{\rm #1}}
\newcommand{\U}{\operatorname{U}}
\newcommand{\GL}{\operatorname{GL}}
\newcommand{\SL}{\operatorname{SL}}
\newcommand{\Ad}{\operatorname{Ad}}
\renewcommand{\mathbf}[1]{{\bf #1}}
\newcommand{\otn}{{\otimes N}}
\newcommand{\otm}{{\otimes M}}
\newcommand{\QED}{{\bf QED}.\\[1ex]}
\newcommand{\piB}{\pi_{\tiny \Box}}

\begin{document}

\title{Optimal Cloning of Pure States, Judging Single Clones }
\author{M. Keyl}
\address{ 
Institut f\"ur Mathematische Physik, TU Braunschweig, 
Mendelssohnstr.3, 38106 Braunschweig, Germany. \\
Electronic Mail: m.keyl@tu-bs.de}
\author{R.~F. Werner}
\address{ 
Institut f\"ur Mathematische Physik, TU Braunschweig, 
Mendelssohnstr.3, 38106 Braunschweig, Germany. \\
Electronic Mail: R.Werner@tu-bs.de}
\date{\today} 

\maketitle 

\begin{abstract}
We consider quantum devices for turning a finite number $N$ of $d$-level 
quantum systems in the same unknown pure state $\sigma$ into $M>N$ 
systems of the same kind, in an approximation of the $M$-fold tensor 
product of the state $\sigma$. In a previous paper it was shown that 
this problem has a unique optimal solution, when the quality of the 
output is judged by arbitrary measurements, involving also the 
correlations between the clones. We show in this paper, that if the 
quality judgement is based solely on measurements of single output 
clones, there is again a unique optimal cloning device, which 
coincides with the one found previously. 
\end{abstract} 


\section{Introduction}
\label{sec:intro}

According to the well known ``no-cloning theorem'' \cite{WooZu} 
perfect copying of quantum information is impossible, i.e. there is 
no machine which takes a quantum system as input and produces two 
systems of the same kind, both of them indistinguishable from the 
input. However, from the point of view of practical applications in 
Quantum Information Theory this Theorem by itself is not very 
useful, because it only asserts that the cloning task cannot be 
performed {\em exactly} --- but then no task can be performed 
exactly by real devices. The fundamental importance of the 
No-Cloning Theorem is expressed much better in stronger versions of 
the Theorem, which also  give explicit lower bounds on the error 
made in any attempt to build a cloning device. Some such bounds have 
been established, as well \cite{Buzek,Hillery}. Even more insight into
the cloning problem is given by results showing how to minimize the
error, i.e., how to construct {\em optimal} cloning devices
\cite{GisMas,Bruss,Bruss2,Buzek2,Zanardi}. Other recent related work
can be found in \cite{GisHut,Gisin,Mor,Cerf1,Cerf2,NiuGriff,Murao}.

In this paper we consider cloning devices, which take as input a 
certain number $N$ of identically prepared systems, and produce a 
larger number $M$ of systems as output. Again, the cloning task is 
to make the output state resemble as much as possible a state of $M$ 
systems all prepared in the same state as the inputs. This variant 
of the problem is of interest as a ``quantum amplifier''. It also 
has a better chance of reasonable success than a cloning device 
operating on single input systems: in the limit of many input 
systems the device can make a good statistical estimate of the input 
density matrix and hence produce arbitrarily good clones. 

Different variants of this problem arise by different choices of the
type of systems and the set of states which should be copied, e.g. 
pure vs.{} mixed states, or a finite number of states arising in a 
cryptographic protocol. In the present paper we are exclusively 
concerned with the cloning of arbitrary unknown pure states. 

A second choice to be made is the precise notion of approximation 
between the output states of the cloning device and the 
(inattainable) target state. Apart from technicalities the basic 
issue here is whether the full states are compared, or only the 
one-clone marginals. Approximation in the first sense means that 
the expectations of all observables, including those testing 
correlations and entanglement between different clones, are close in 
the two states being compared. On the other hand, 
approximation in the second sense means closeness of expectations of 
single clone observables only. Perhaps this second condition has 
more of the flavour of the No-Cloning Theorem, since in that 
theorem, too, the requirement is that each single (!) clone be 
indistinguishable from the input.

In \cite{part1} we showed that the pure state cloning problem with 
all-particle test criterion has a unique optimal solution. In this 
paper we show the same for the single-particle test criterion, and 
that the two optimal cloning devices are actually the same. The 
difference between the two results may not seem great. However, the 
result in the present paper required much heavier mathematical 
machinery, and we believe it to be considerably deeper. The reason 
is that one-particle tests by far do not exhaust the linear space of 
$M$-particle observables. In particular, all correlations of the 
cloner's output are ignored by the test, which would make a {\em 
unique} optimal solution appear rather unlikely. Nevertheless, this 
is what we prove. 

\section{Statement of the Problem and Main Result}
\label{sec:problem}

Let us start with a precise formulation of the question we are going 
to consider. First of all, we will study throughout this paper only 
$d$-level systems with arbitrary but finite $d$. Hence the one 
particle Hilbert space ${\cal H}$ we are using is ${\cal 
H}=\Bbb{C}^d$. The Hilbert space for the input to the cloning device 
is therefore the $N$-fold tensor product ${\cal H}^{\otimes N}$ of 
${\cal H}$ with itself. In fact, because we only consider tensor 
powers of pure states as inputs, it suffices to take the subspace of 
${\cal H}^{\otimes N}$ spanned by vectors of the form $\phi^{\otimes 
N}$ with $\phi\in\H$. This is precisely the ``Bose'' subspace ${\cal 
H}^{\otimes N}_+\subset{\cal H}^{\otimes N}$, i.e., the space of 
vectors invariant under all permutations. The output Hilbert space 
will be ${\cal H}^{\otimes M}$ with $M>N$. On this space we cannot 
impose an a priori symmetry restriction, although such a restriction  
will come out a posteriori, as a special property of optimal cloning 
devices. 

A {\em cloning map} is a completely positive, unital map $T: {\cal 
B}({\cal H}^\otm) \to {\cal B}({\cal H}^\otn_+)$. This describes the 
action of the device on observables. Its (pre-)dual, describing the 
same operation in terms of states, will be denoted%
\footnote{
In \cite{part1} the symbol $T$ was used for $T_*$. In contrast to 
\cite{part1} the key arguments in the present paper are phrased more 
readily in terms of the map on observables than in terms of the map 
on states. Therefore, we decided to change this notation, which then 
also agrees better with the usage for completely positive maps on 
operator algebras.}
by $T_*: {\cal B}_*({\cal H}^\otn_+) \to {\cal B}_*({\cal H}^\otm)$.  
If we identify states with density operators, this means that 
$\tr\bigl(\rho\,T(A)\bigr)=\tr\bigl(T_*(\rho) A\bigr)$ for arbitrary 
density operators $\rho$ and observables $A$. The input of the 
cloning device are $N$ systems, prepared independently according to 
the same state $\sigma$. Thus the overall input state is 
$\sigma^\otn$. We will assume $\sigma$ to be pure, i.e., the density 
matrix of $\sigma$ is a one-dimensional projection onto a wave 
vector $\psi\in\H$, say. Then $\sigma^{\otimes N}$ is the projection 
onto the vector $\psi^{\otimes N}$. The output of the cloning device 
is the state $T_*(\sigma^\otn)$, which is a (generally 
entangled) state of $M>N$ systems. Our aim is to design $T$ so that 
the output states $T_*(\sigma^\otn)$ approximate the product states 
$\sigma^\otm$.

The one particle observables, on which the comparison will be based, 
will be written as 
$a_{(k)} 
   = \Bbb{1}^{\otimes({k-1})} \otimes a \otimes 
      \Bbb{1}^{\otimes({M-k})}
     \in {\cal B}({\cal H}^\otm)$,  
for all $a \in {\cal B}({\cal H})$. 
Thus the optimal cloning problem for arbitrary pure 
input states $T$ is to make the expectations
\begin{eqnarray*}
  \tr(a_{(k)} T_*(\sigma^\otn)) 
      &=& \tr(T(a_{(k)})\sigma^\otn) 
       =  \langle \psi^\otn,  T(a_{(k)}) \psi^\otn\rangle \cr
\text{and}\qquad
   \tr(a_{(k)} \sigma^\otm)
      &=& \tr(a \sigma)
       =  \langle \psi, a \psi \rangle 
\end{eqnarray*}
as similar as possible for arbitrary one-particle observables $a$ 
and one-particle vectors $\psi$. Of course, when taking a supremum 
over such differences, the size of $a$ has to be constrained 
somehow. We will choose the constraint $0\leq a\leq\idty$, which is 
to say that the above two expressions have an immediate 
interpretation as probabilities. The largest difference of such 
probabilities is now the error functional for cloning maps, which we 
will seek to minimize:                                     
\begin{equation}
  \Label{eq:DeltaOne}
  \Delta_{\rm one}(T) 
     = \sup_{a,\psi,k} \left| \langle \psi^\otn, T(a_{(k)}) 
             \psi^\otn\rangle - \langle \psi, a \psi \rangle \right|
\end{equation}
where the supremum is taken over all $\psi \in {\cal H}$ with 
$\|\psi\|=1$, all operators $a \in {\cal B}({\cal H})$ with $0\leq 
a\leq\idty$, and all integers $1 \leq k \leq M$. 

The corresponding quantity based on tests of the full output state 
(including correlations) is 
\begin{displaymath}
   \Delta_{\rm all}(T) = \sup_{A}\sup_{\sigma, {\rm pure}} \left|
  \tr\left( T(A) \sigma^\otn \right) - \tr(A \sigma^\otm) \right|,
\end{displaymath}
where the supremum is taken over all $A \in {\cal B}({\cal H}^\otm)$
with $0\leq A\leq\idty$ and over all pure states $\sigma \in
{\cal B}_*({\cal H})$. Due to the properties of the trace norm
$\|\,\cdot\,\|_1$ this functional can be expressed by
\begin{equation} \Label{eq:DeltaAll}
  \Delta_{\rm all}(T) 
     = \sup_{\sigma, {\rm pure}} \|T_*(\sigma^\otn) -
            \sigma^\otm\|_1.
\end{equation}
It turns out that there is exactly one cloning map $\widehat{T}$ which
minimizes this error functional. This can be proven with a minor 
adaptation of the arguments in \cite{part1}, which start from a 
slightly different criterion, namely the maximization of the 
``fidelity'' 
${\cal  F}(T_*) = \sup_{\sigma, {\rm pure}}
     (\sigma^\otm T_*(\sigma^\otn))$. 
The unique solution $T=\widehat T$ minimizing (\ref{eq:DeltaAll}), or 
maximizing ${\cal  F}(T_*)$, is best expressed in terms of its 
action on states, i.e.,  
\begin{equation}\label{bestT}
  \widehat{T}_*(\rho) = \frac{d[N]}{d[M]} S_M(\rho\otimes 
  \Bbb{1}^{M-N})S_M.
\end{equation}
Here $d[N] = {d + N -1 \choose N}$ denotes the dimension of the
symmetric subspace ${\cal H}_+^ \otn$,  $S_M$ is the projection from
${\cal H}^\otm$ to ${\cal H}_+^\otm$, and $\rho$ is an arbitrary 
density operator on ${\cal H}_+^\otn$. In \cite{part1} we also 
computed the one-site restriction of the output states of this 
cloner:  
\begin{displaymath}
    \tr\left( \widehat T(a_{k}) \sigma^\otn \right)
        = \gamma(\widehat T) \sigma(a)
          + (1-\gamma(\widehat T))\tr(a)/d
\quad,
\end{displaymath}
where 
\begin{displaymath}
    \gamma(\widehat T)={N\over N+d}\, {M+d\over M}
\end{displaymath}
is the so-called Black Cow factor of $\widehat T$, interpreted as a 
``shrinking factor of the Poincar\'e sphere'' in the discussions of 
the qubit ($d=2)$ case. This makes it easy to verify the case of 
equality in the following Theorem, which is our main result.

\proclaim\XTheorem\ Theorem.
For any cloning map $T: {\cal B}({\cal H}^\otm) \to {\cal B}({\cal
  H}^\otn_+)$ we have
\begin{displaymath}
  \Delta_{\rm one}(T) \geq \frac{d-1}{d} \left| 1 - \frac{N}{N+d}
  \frac{M+d}{M} \right|
\end{displaymath}
with equality iff $T=\widehat{T}$ with $\widehat{T}$ from 
equation (\ref{bestT}).

\section{Finding the optimal cloning map}
\label{sec:finding}

\subsection{Reduction to the covariant case}
\label{sec:reduction}

In this section we will give the proof of our main theorem, apart 
from some group theoretical Lemmas, which will be proved in 
Appendix~A. Throughout, the symmetry of sitewise unitary rotation of 
clones and input states will play a crucial role. The necessary 
background information on unitary representations of ${\rm SU}(d)$ 
will also be supplied in Appendix~A.

We establish some notation first. By ${\rm U}(d)$ we will denote the 
group of unitary $d\times d$-matrices, i.e., the unitary group on 
our underlying one-particle space $\H\equiv\Cx^d$. Unitary 
representations of this group will be denoted by the letter $\pi$ 
with suitable indices. $\piB$ is the defining representation on 
$\Cx^d$, and its $n^{\rm th}$ tensor power, acting on 
$\H^{\otimes N}$  by the operators 
$\piB^{\otimes N}(u)=u^{\otimes N}$ is $\piB^{\otimes N}$. 
The restriction of this representation 
to the symmetric subspace $\H^{\otimes N}_+$ will be denoted by 
$\pi_N^+$. Thus a cloning map 
$T:\B(\H^{\otimes M})\to\B(\H^{\otimes N}_+)$ is called 
${\rm U}(d)$-{\em covariant}, if 
\begin{equation}
T\bigl(\piB^{\otimes M}(u)A\piB^{\otimes M}(u)^*\bigr)
      =\pi_N^+(u)T(A)\pi_N^+(u)^*
\end{equation}
This equation merely expresses that $T$ does not prefer any 
direction in $\H$. It would be a natural initial assumption for good 
cloning devices but, of course, in our case it will come out as a 
result of the minimization: $\widehat T$ from equation (\ref{bestT}) 
is obviously covariant, because $S_M$ commutes with all 
$\piB^{\otimes M}(u)$. It is convenient to state the covariance 
condition as a fixed point property: we define the action $\tau$ of 
unitary rotations on cloning maps by 
\begin{equation}
    (\tau_uT)(A)=\pi_N^+(u)^* T\Bigl(\piB^{\otimes M}(u)
                  A\piB^{\otimes M}(u)^*\Bigr)\pi_N^+(u),
\end{equation}
so that $T$ is covariant iff $\tau_u(T)=T$ for all $u\in{\rm U}(d)$. 
We denote by $\overline{T}$ the average of $\tau_uT$ with respect to 
$u$, i.e., 
\begin{equation}
    \overline{T}=\int du\ \tau_u(T),
\Label{mean}
\end{equation}
where ``$du$'' denotes the normalized Haar measure on $\U(d)$. 

The fact that the cloning error $\Delta_{\rm one}$ does not single 
out a direction on $\H$ either is expressed by the --- easily 
verified --- equation 
\begin{equation}
    \Delta_{\rm one}(\tau_uT)=\Delta_{\rm one}(T).
\end{equation}
Similarly, we can get an estimate of 
$\Delta_{\rm one}(\overline{T})$: 
The functional $\Delta_{\rm one}$ is defined as the supremum of a 
set of convex expressions in $T$. Therefore, it is convex, and 
$\Delta_{\rm one}(\overline{T})\leq \Delta_{\rm one}(T)$. So as long 
as we are only interested in finding {\em some} cloning map with 
minimal $\Delta_{\rm one}$, we may restrict attention to covariant 
ones. 

There is a similar simplification, which we can make ``without loss 
of cloning quality'': $\Delta_{\rm one}$ is invariant under a change 
of the ordering of the clones. That is to say, if 
$V:\H^{\otimes M}\to\H^{\otimes M}$ is a permutation operator, and 
if we define $\tau_VT$ by $(\tau_VT)(A)=T(VAV^*)$, we may replace 
$T$ by its average over permutations without loss of cloning 
quality. That is, we may assume that $\tau_VT=T$ for all 
permutations $V$. We will refer to this property as {\em permutation 
invariance}. 

Our strategy is now to assume ${\rm U}(d)$-covariance and 
permutation invariance of $T$, and to show that there is a unique 
solution to the variational problem with these additional 
properties. The above convexity argument then implies that no other 
cloning map can do better. But since the functional 
$\Delta_{\rm one}$ is not strictly convex, we will need an extra 
step to establish uniqueness. This we will do in 
subsection~\ref{sec:unique} by showing that any cloning map whose 
mean is the optimal covariant cloner has to be covariant itself.

\subsection{Reduction to the extremal covariant case}

The functional $\Delta_{\rm one}$ involves only operators 
$T(A)$ with $A$ of the special form $A=a_{(k)} 
   = \Bbb{1}^{\otimes({k-1})} \otimes a \otimes 
      \Bbb{1}^{\otimes({M-k})}
     \in {\cal B}({\cal H}^\otm)$.
Now due to permutation invariance $T(a_{(k)})$ does not depend on $k$,
and we have 
\begin{equation}\Label{e.Tak}
    T(a_{(k)})=\frac1M T\Bigl(\sum_k a_{(k)}\Bigr).
\end{equation}
What makes this equation useful is that on the right hand side $T$ 
is now applied to one of the generators of the representation 
$\piB^{\otimes N}$: we have 
$\exp\bigl(i\sum_{k=1}^Ma_{(k)}\bigr)
    =\bigl(\exp(ia)\bigr)^{\otimes M}$. 
Because $T$ is covariant, we 
can determine how the operators in equation~(\ref{e.Tak}) transform 
under ${\rm U}(d)$-rotations:
\begin{equation}
\label{transTgen} 
   \pi_N^+(u)T(a_{(k)})\pi_N^+(u)^*
     =T\bigl((uau^*)_{(k)}\bigr),
\end{equation}
where the multiplication of $a$ and $u$ on the right hand side is in 
the $d\times d$-matrices. This property fixes the ``transformation 
behaviour'' of the operators $T(a_{(k)})$, and as we will see, this 
essentially fixes the tuple of operators $T(a_{(k)})$. 
Of course, $a=\idty$ in (\ref{transTgen}) simply leads to 
$T(\idty_{(k)})=T(\idty)=\idty$. The operator $i\Bbb{1}$ is the
(anti-hermitian) generator of the subgroup of unitaries multiplying
each vector with the same phase. More interesting are the generators
of ${\rm SU}(d)$, in which such trivial phases have been
eliminated. These generators, in other words the Lie algebra
$\Lie{SU}(d)$, are exactly the traceless anti-hermitian $d\times
d$-matrices. In the qubit case ($d=2$) $\Lie{SU}(2)$ is spanned by the
Pauli matrices (multiplied by $i$), and $3$-tuples of operators
transforming like the generators are known in physics literature as
``vector operators''. It is well-known that, due to the simple
reducibility of ${\rm SU}(2)$, each irreducible representation of
${\rm SU}(2)$ contains exactly one vector operator (up to a factor),
namely the generators (angular momentum operators) of the
representation themselves. So all operators $T(a_{(k)})$ are
determined by the single numerical factor relating the operators
$T(a_{(k)})$ to the generators of the irreducible representation
$\pi_N^+$.  

It turns out that the same idea works in the ${\rm SU}(d)$-case for 
arbitrary $d$. In order to state it precisely, we need a notation 
for the Lie algebra representation associated with a unitary 
representation of a Lie group. We define $\partial\pi(X)$ to be the 
anti-hermitian generator of the one-parameter subgroup generated by
$X$, i.e., 
\begin{equation}
   \partial\pi(X)
      =\left.\frac{d}{dt}\pi\bigl(e^{tX}\bigr)
       \right\vert_{t=0}. 
\end{equation}
Then the desired property of a representation is stated in the 
following definition: 

\proclaim \XDnondeg\ Definition. 
Let $\pi:G\to\B(\H_\pi)$ be a finite dimensional unitary 
representation of a Lie group $G$ with Lie algebra $\Lie G$. Then 
$\Lie G$ is said to be {\bf non-degenerate} in $\B(\H_\pi)$  with 
respect to $\pi$, if any linear operator 
$L:\Lie G\to\B(\H_\pi)$ with the covariance property 
$\pi(g)L(X)\pi(g)^*=L(gXg^{-1})$ is of the form 
$L(X)=\lambda \partial\pi(X)$, for some factor $\lambda\in\Cx$.

As we argued above, $\Lie{SU}(2)$ is non-degenerate in {\em every} 
irreducible representation of ${\rm SU}(2)$. However, for $d\geq3$ 
we can find representations containing degenerate copies of the 
generators, and we have to make sure that the special 
representations occurring in the present problem are of the ``good'' 
kind. This is the content of the following Lemma, proved in 
Appendix~A. 

\proclaim\XLnondeg\ Lemma. 
$\Lie{SU}(d)$  is non-degenerate in $\B(\H^{\otimes N}_+)$  with 
respect to $\pi_N^+$. 

\proclaim\Xomega\ Corollary. 
Let $\pi:{\rm U}(d)\to\B(\H_\pi)$ be a unitary representation, and 
let $T:\B(\H_\pi)\to\B(\H^{\otimes N}_+)$ be a completely positive 
normalized and ${\rm U}(d)$-covariant map, i.e. $T(\pi(u)A\pi(u)^*) =
\pi_N^+(u)T(A)\pi_N^+(u)^*$. Then there is a number $\omega(T)$ such
that  
\begin{displaymath}
   T(\partial\pi(a))=\omega(T) \sum_{k=1}^N a_{(k)},
\end{displaymath}
for every $a\in\B(\H)$ with $\tr(a)=0$. 

Given $\omega(T)$ for $\pi=\piB^{\otimes M}$, we can compute the 
cloning error  $\Delta_{\rm one}(T)$ as follows: given $a\in\B(\H)$ 
with $0\leq a\leq\idty$, we can write $a=\alpha\idty+a'$ with 
${\rm tr}a'=0$. Then 
\begin{eqnarray*}
   T(a_{(k)})&=&\alpha\idty
            +\frac{1}{M}T\Bigl(\sum_{l=1}^Ma'_{(l)}\Bigr) \\
        &=&\alpha\idty
           +\frac{\omega(T)}{M} \Bigl(\sum_{l=1}^Na'_{(l)}\Bigr)
\end{eqnarray*}
and with $a'=a-\alpha\Bbb{1}$ and $\alpha = \frac{\tr a}{d}$
\begin{displaymath}
     T(a_{(k)})=\frac{\tr a}{d}\left(1-\frac{N\omega(T)}{M}\right)\Bbb{1}
     +\frac{\omega(T)}{M}\left(\sum_{l=1}^Na_{(l)}\right).
\end{displaymath}
In any state $\psi\in\H$ we get 
\begin{displaymath}
    \langle\psi, a\psi\rangle 
      -\langle\psi^{\otimes N}, T(a_{(k)})
              \psi^{\otimes N}\rangle
    = (1 - \gamma(T))\left(\langle \psi,a\psi\rangle-
      \frac{\tr a}{d}\right)
\end{displaymath}
where $\gamma(T) = \frac{N}{M} \omega(T)$ is the Black-Cow factor
already mentioned in Section \ref{sec:problem}. With
\begin{displaymath}
  \sup_{\psi,a} \left(\langle \psi,a\psi\rangle- \frac{\tr
  a}{d}\right)=\frac{d-1}{d}   
\end{displaymath}
we get
\begin{equation}
    \Delta_{\rm one}(T)
       =\frac{d-1}{d}\left\vert 1-\frac{N}{M}\omega(T)\right\vert =
       \frac{d-1}{d} \left| 1- \gamma(T)\right|.
\Label{Delomega}
\end{equation}
We remark that the largest possible $\omega(T)$, to be determined 
below, still makes the second term in the absolute value less than 
$1$, so we could omit the absolute value signs. In any case, we will 
only seek to maximize $\omega(T)$ from now on, ignoring the 
possibility of $\omega(T)>M/N$, anticipating that it will be ruled 
out by the result of the maximization anyway. 

An important observation about the Corollary and 
formula~(\ref{Delomega}) is that $\omega$ is clearly an {\em affine} 
functional on the convex set of covariant cloning maps (i.e., 
$\omega$ respects convex combinations). Whereas we previously used 
the convexity of $\Delta_{{\rm one}}$ to conclude that averaging over 
rotations and permutations (and hence a move towards the interior of 
the convex set of cloning maps) generally improves the cloning 
quality, we now see that the optimum can be sought, as for any 
affine functional, on the extreme boundary of the subset of 
covariant cloning maps. Therefore our next steps will be aimed at 
the determination of the extremal ${\rm U}(d)$-covariant and 
permutation invariant cloning maps, and, subsequently the solution 
of the variational problem for these extremal cases. 

\subsection{Convex decomposition of covariant cloning maps}
\label{sec:covex-decompo}

For the first reduction step we use the close connection between the 
permutation operators on $\H^{\otimes M}$ and the representation 
$\piB^{\otimes M}$. Let $\bigl(\piB^{\otimes M}\bigr)'$ denote the 
algebra of all operators on $\H^{\otimes M}$ commuting with all 
$\piB^{\otimes M}(u)\equiv u^{\otimes M}$. This algebra consists 
precisely of the linear combinations of permutation unitaries 
\cite[Theorem IX.11.5]{Simon}. So consider a reduction of
$\piB^{\otimes M}$ into irreducibles, i.e., an orthogonal
decomposition of the identity into minimal projections
$E_\alpha\in(\piB^{\otimes M})'$. Then due to covariance the operators
$T(E_\alpha)$ commute with all $\pi_N^+(u)$, and because the latter
representation is irreducible, they must be multiples of the identity,
$T(E_\alpha)=r_\alpha\idty$, say. Because $T(VA)=T(AV)$ for
permutation operators $V$, we also have $T(AE_\alpha)=T(E_\alpha
AE_\alpha)$. Hence  
\begin{equation}
   T_\alpha(A)=r_\alpha^{-1}T(E_\alpha AE_\alpha)
\end{equation}
is a legitimate cloning map in its own right (provided 
$r_\alpha\neq0$). Moreover, 
\begin{equation}
  T(A)=\sum_\alpha T(AE_\alpha)
      =\sum_\alpha T(E_\alpha AE_\alpha)
      =\sum_\alpha r_\alpha T_\alpha(A)
\end{equation}
is a convex decomposition of the given $T$ into such summands. 
Maximizing $\omega(T)=\sum_\alpha r_\alpha\omega(T_\alpha)$ thus 
means concentrating the coefficients $r_\alpha$ on those $\alpha$, 
for which $\omega(T_\alpha)$ is maximal. At this stage it is perhaps 
already plausible that only the summand $T_\alpha$, for which 
$E_\alpha=S_M$ is the projection onto the symmetric subspace will 
give the best $\omega(T_\alpha)$, because this is the space 
supporting the pure states $\sigma^{\otimes M}$ the cloner is 
supposed to approximate. In fact, for the optimization of 
$\Delta_{\rm all}$ in \cite{part1} this idea leads directly to a 
simple solution. In the present case we found no direct proof of 
this plausible statement. 

We therefore have to enter into the further convex decomposition of 
each $T_\alpha$. The output states of this cloning map are supported 
by $\H_\alpha\equiv E_\alpha\H^{\otimes M}$, and we will restrict 
$T_\alpha$ accordingly, i.e., we consider it as a covariant map 
$T_\alpha:\B(\H_\alpha)\to\B(\H^{\otimes N}_+)$, which is covariant 
with respect to the restricted representation 
$\pi_\alpha=\pi^{\otimes M}\rstr\H_\alpha$ and $\pi_N^+$.

As for any completely positive map, the convex decompositions of 
$T_\alpha$ are governed by the Stinespring dilation 
\cite{Stinespring}. Since we are looking, more specifically, for 
decompositions into covariant completely positive maps, we have to 
invoke a ``covariant'' version of the Stinespring dilation Theorem 
\cite{Scutaru}, which is stated in Appendix~B for the convenience of 
the reader. According to this Theorem we can write a covariant 
completely positive 
$T_\alpha:\B(\H_\alpha)\to\B(\H^{\otimes N}_+)$ as 
$T_\alpha(A)=V^*(A\otimes\idty_\K)V$, where $\K$ is some auxiliary 
Hilbert space carrying a unitary representation 
$\widetilde\pi:{\rm U}(d)\to\B(\K)$, and 
$V:\H^{\otimes N}_+\to\H_\alpha\otimes\K$ is an isometry intertwining 
the respective representations, i.e., 
\begin{equation}
   V\pi_N^+(u)=\bigl(\pi_\alpha(u)\otimes\widetilde\pi(u)\bigr)V. 
\end{equation}
The convex reduction theory of $T_\alpha$ is now the same as the 
reduction theory of $\widetilde\pi$ into irreducibles: if $F_\beta$ 
is a minimal projection in the algebra $\widetilde\pi'$, and hence 
$\widetilde\pi\rstr F_\beta\K$ is irreducible, then 
$A\mapsto V^*(A\otimes F_\beta)V$ is a covariant map, which cannot 
be further decomposed into a sum of covariant completely positive 
maps (see Appendix~B). Note that $V^*(\idty\otimes F_\beta)V$ 
commutes with the irreducible representation $\pi_N^+$, so that once 
again this summand is normalized up to a factor: 
$V^*(\idty\otimes F_\beta)V=r_{\beta}\idty$.  Therefore 
$T_\alpha=\sum_\beta r_\beta T_{\alpha\beta}$, where each 
$T_{\alpha\beta}(A)=r_\beta^{-1}V^*(A\otimes F_\beta)V$ is again an 
admissible cloning map. The following statement summarizes the 
result of the decomposition theory of $T$. 

\proclaim\Xdecompo\ Proposition. 
Let $T:\B(\H^{\otimes M})\to\B(\H^{\otimes N}_+)$ be a 
${\rm U}(d)$-covariant and permutation invariant cloning map. 
Then $T$ is a convex combination 
$T=\sum_{\alpha\beta}r_{\alpha\beta}T_{\alpha\beta}$ such that 
each $T_{\alpha\beta}$ is of the following special form:  
$T_{\alpha\beta}(A)=V^*(A\otimes\idty_\beta)V$,  
where $V$ is an intertwining isometry between $\pi_N^+$ and 
$\pi_\alpha\otimes\pi_\beta$, such that 
$\pi_\alpha:{\rm U}(d)\to\B(\H_\alpha)$ is an irreducible 
subrepresentation of $\piB^{\otimes M}$, and 
$\pi_\beta:{\rm U}(d)\to\B(\H_\beta)$ is also an irreducible unitary 
representation. 

This Proposition summarizes all that is needed for the further 
treatment of the variational problem. However, we could have made a 
slightly stronger statement by eliminating the non-uniqueness 
introduced by the choice of the minimal projections $E_\alpha$. If 
the subrepresentations $\pi_\alpha$ and $\pi_{\alpha'}$ are 
unitarily equivalent, then they can be connected by a unitary, which 
is again a linear combination of permutations. Hence the contribution
of the term $r_\alpha T_\alpha=\sum_\beta
r_{\alpha\beta}T_{\alpha\beta}$ to $\omega(T)$ depends only on the
isomorphism type of $\pi_\alpha$. 

What we cannot assert in general, however, is that $V$ is determined 
by the isomorphism types of $\pi_\alpha$ and $\pi_\beta$: among the 
groups ${\rm SU}(d)$ only $d=2$ is ``simply reducible'', which means 
that the space of intertwiners between $\pi_\gamma$ and 
$\pi_\alpha\otimes\pi_\beta$ is at most one-dimensional for 
arbitrary irreducible representations 
$\pi_\alpha,\pi_\beta,\pi_\gamma$. In the following subsection we 
will therefore focus on the qubit case, and show 
how to determine $\omega(T_{\alpha\beta})$ from the representations 
involved. This procedure will then be generalized to arbitrary $d$, 
and it will turn out that, perhaps surprisingly, in the general case 
$\omega(T_{\alpha\beta})$ also depends on $\pi_\alpha,\pi_\beta$ 
only up to unitary equivalence. 

\subsection{Maximizing $\omega$ in the case $d=2$}
\label{sec:maxOmega-su2}

For $d=2$ the representations of ${\rm SU}(2)$ are conventionally 
labelled by their ``total angular momentum'' $j=0,1/2,1,\ldots$. The 
irreducible representation $\pi_j$ has dimension $2j+1$, and is 
isomorphic to $\pi_{N}^+$ with $N=2j$ in the notation used above. 
For $j=1$ we get the $3$-dimensional representation isomorphic to the
rotation group, which is responsible for the importance of this group
in physics. In a suitable basis $X_1,X_2,X_3$ of the Lie algebra
$\Lie{SU}(2)$ we get the commutation relations $\lbrack
X_1,X_2\rbrack=X_3$, and cyclic permutations of the indices
thereof. In the $j=1$ representation $\partial\pi_1(X_k)$ generates
the rotations around the $k$-axis in $3$-space. The Casimir operator
of ${\rm SU}(2)$ is the square of this vector operator, i.e.,
$\widetilde{\bf C}_2=\sum_{k=1}^3X_k^2$. In the representation $\pi_j$
it is the scalar $j(j+1)$, i.e., if we extend the representation
$\partial\pi$ of the Lie algebra to the universal enveloping algebra
(which also contains polynomials in the generators), we get
$\partial\pi_j(\widetilde{\bf C}_2)=j(j+1)\idty$. We can use this to
determine $\omega(T_{\alpha\beta})$ for arbitrary irreducible 
representations. This computation can be seen as an elementary
computation of a so-called $6j$-symbol (see also \cite{FNW}  for a 
context in which the same computation arises), but we will not need to  
invoke any of the $6j$-machinery.

So let $V$ be an intertwining isometry between $\pi_\gamma$ and 
$\pi_\alpha\otimes\pi_\beta$, where 
$\alpha,\beta,\gamma\in\lbrace0,1/2,\ldots\rbrace$ label 
irreducible representations. Then $\omega$ is defined by 
\begin{equation}
\Label{eq:omega-def}
   \omega\cdot \partial\pi_\gamma(X_k)
   =V^*(\partial\pi_\alpha(X_k)\otimes\idty_\beta)V. 
\end{equation}
We multiply this equation by $\partial\pi_\gamma(X_k)$, use the 
intertwining property of $V$ in the form 
$V\partial\pi_\gamma(X)
   =\bigl(\partial\pi_\alpha(X)\otimes\idty_\beta
     +\idty_\alpha\otimes \partial\pi_\beta(X)\bigr)V$, 
and sum over $k$ to get 
\begin{equation}
  \Label{eq:omega-casimir}
   \omega\cdot \partial\pi_\gamma(\widetilde{\bf C}_2)
   =V^*\bigl(\partial\pi_\alpha(\widetilde{\bf
   C}_2)\otimes\idty_\beta\bigr)V + \sum_k
   V^*\bigl(\partial\pi_\alpha(X_k)\otimes
   \partial\pi_\beta(X_k)\bigr)V.  
\end{equation}
The tensor product in the second summand can be re-expressed in 
terms of Casimir operators as 
\begin{displaymath}
\sum_k \bigl(\partial\pi_\alpha(X_k)\otimes \partial\pi_\beta(X_k)\bigr)
   =\frac12\sum_k \bigl(\partial\pi_\alpha(X_k)\otimes\idty_\beta
                     +\idty_\alpha\otimes \partial\pi_\beta(X_k)\bigr)^2
    -\frac12 \partial\pi_\alpha(\widetilde{\bf C}_2)\otimes\idty_\beta
    -\frac12 \idty_\alpha\otimes \partial\pi_\alpha(\widetilde{\bf C}_2).
\end{displaymath}
Inserting this into the previous equation, using the intertwining 
property once again, and inserting the appropriate scalars for 
$\partial\pi(\widetilde{\bf C}_2)\equiv \widetilde{C}_2(\pi)\idty$, 
we find that 
$\omega\cdot \widetilde{C}_2(\pi_\gamma)=
\widetilde{C}_2(\pi_\alpha)+\frac12\bigl(
\widetilde{C}_2(\pi_\gamma)-\widetilde{C}_2(\pi_\alpha)-
\widetilde{C}_2(\pi_\beta)\bigr)$, 
and hence 
\begin{equation}\Label{omegaC2}
   \omega=\frac12+ 
   \frac{\widetilde{C}_2(\pi_\alpha) -
     \widetilde{C}_2(\pi_\beta)}{2\widetilde{C}_2(\pi_\gamma)}. 
\end{equation}
Note that we have only used the fact that the Casimir operator ${\bf 
\widetilde{C}_2}$ is some fixed quadratic expression in the
generators. This is also true for ${\rm SU}(d)$. Hence
equation~(\ref{omegaC2}) also holds in the general case. In
particular, we have shown that for the purpose of optimizing
$\omega(T_{\alpha\beta})$ only the isomorphism types of $\pi_\alpha$
and $\pi_\beta$ are relevant, but not the particular intertwiner $V$.

Specializing again to the case $d=2$, we find 
\begin{equation}\Label{omegaSU2}
   \omega=\frac12+ 
           \frac{\alpha(\alpha+1)-\beta(\beta+1)}{2\gamma(\gamma+1)}.
\end{equation}
Here $\gamma=N/2$ is fixed by the number $N$ of input systems.  
$\alpha$ is constrained by the condition that $\pi_\alpha$ 
must be a subrepresentation of $\pi_{j=1/2}^{\otimes M}$, which is 
equivalent to $\alpha\leq M/2$. Finally, $\beta$ is constrained by 
the condition that there must be a non-zero intertwiner between 
$\pi_\gamma$ and $\pi_\alpha\otimes\pi_\beta$. It is well-known that 
this condition is equivalent to the inequality 
$\vert{\alpha-\beta}\vert\leq\gamma\leq{\alpha+\beta}$. This is the 
same as the ``triangle inequality'': the sum of any two of 
$\alpha,\beta,\gamma$ is larger than the third. The area of 
admissible pairs $(\alpha,\beta)$ is represented in Fig.~1. 

Since $x\mapsto x(x+1)$ is increasing for $x\geq0$, we maximize 
$\omega$ with respect to $\beta$ in equation~(\ref{omegaSU2}) if we 
choose $\beta$ as small as possible, i.e., 
$\beta=\vert{\alpha-\gamma}\vert$. 
Then the numerator in equation~(\ref{omegaSU2}) becomes 
\begin{displaymath}
  \alpha(\alpha+1)-\beta(\beta+1)
    =2\alpha\gamma-\gamma^2 
      +\max\lbrace{\gamma,2\alpha-\gamma}\rbrace, 
\end{displaymath}
which is strictly increasing in $\alpha$. Hence the maximum 
\begin{equation}
   \omega_{\rm max}=\frac{M+2}{N+2}
\end{equation}
is attained for and only for $\alpha=M/2$ and $\beta=(M-N)/2$. 

\begin{figure}[htbp]
  \begin{center}
    \unitlength0.05mm
    \linethickness{0.5pt}
    \begin{picture}(1646,1344)
      \put(156,9){\vector(0,1){1320}}
      \put(76,89){\vector(1,0){1510}}
      \put(156,289){\line(1,1){960}}
      \put(156,289){\line(1,-1){200}}
      \put(356,89){\line(1,1){760}}
      \put(1116,849){\line(0,1){400}}
      \put(1116,849){\circle*{20}}
      \put(80,1309){$\beta$}
      \put(1516,9){$\alpha$}
      \put(356,109){\line(0,-1){40}}
      \put(176,289){\line(-1,0){40}}
      \put(176,849){\line(-1,0){40}}
      \put(1116,109){\line(0,-1){40}}
      \put(1176,829){$\omega_{\rm max}$}
      \put(316,9){$N/2$}
      \put(1076,9){$M/2$}
      \put(-20,269){$N/2$}
      \put(-240,829){$(M-N)/2$}
    \end{picture}
    \caption{Area of admissible pairs $(\alpha,\beta)$.}
    \label{fig:fig1}
  \end{center}
\end{figure}
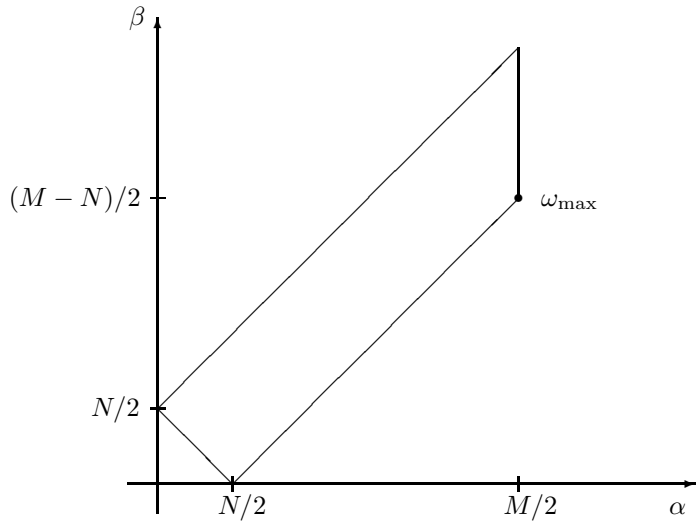

Note that the seemingly simpler procedure of first maximizing 
$\alpha$ and then minimizing $\beta$ to the smallest value 
consistent with $\alpha=M/2$ leads to the same result, but is 
fallacious because it fails to rule out possibly larger values of
$\omega$ in the lower triangle of the admissible region in Fig.~1.  
The same problem arises for higher $d$, and one has to be careful to 
find a maximization procedure which takes into account all 
constraints. 

\subsection{Maximizing $\omega$ in the general case}
\label{sec:maxOmega-general}

Let us generalize now the previous discussion to arbitrary but finite
$d$. In this case irreducible representations of $\U(d)$ are labelled,
according to Section \ref{sec:unit-groups:reps} by their highest weight
${\bf m}=(m_1, \dots, m_d)$. Hence we can decompose $T: {\cal B}({\cal
  H}^\otm) \to {\cal B}({\cal H}^\otn_+)$ as described in
Prop. \Xdecompo\  into the sum $T = \sum_{({\bf m},{\bf n}) \in W}
r_{{\bf m},{\bf n}} T_{{\bf m},{\bf n}}$, taken over the set 
\begin{displaymath}
  W = \{ (\mathbf{m},\mathbf{n}) \in \Bbb{Z}^d_+ \times \Bbb{Z}^d_+ \, |
  \, \pi_\mathbf{m} \subset \piB^\otm \mbox { and } \pi_N^+ \subset
  \pi_\mathbf{m} \otimes \pi_\mathbf{n} \}.
\end{displaymath}
Here $\Bbb{Z}^d_+$ is an abbreviation for the set of all possible
weights of irreducible $\U(d)$ representations, i.e. $\Bbb{Z}^d_+ = \{
(m_1, \dots, m_d) \, | \, m_1 \geq m_2 \geq \dots \geq m_d \}$.

Our task is now to determine $(\mathbf{m},\mathbf{n}) \in W$ such that
$\omega = \omega(T_{\mathbf{m},\mathbf{n}})$ becomes maximal. To this
end we consider in analogy to (\ref{eq:omega-def}) the equation 
\begin{equation}
\Label{eq:omega-def-2}
   \omega\cdot \partial\pi_N^+(X) =
   V^*(\partial\pi_\mathbf{m}(X)\otimes\idty_\mathbf{n})V, \quad
   \forall X \in \goth{su}(d)
\end{equation}
where $V$ is an intertwining isometry between $\pi_N^+$ and
$\pi_\mathbf{m} \otimes \pi_\mathbf{n}$.
Note that equation (\ref{eq:omega-def-2}) is valid only for $X \in
\goth{su}(d)$ (and not for $X \in \goth{u}(d)$ in general). Hence we
have to consider the second order Casimir operator $\widetilde{\bf
  C}_2$ of $\SU d$ which is given, according to 
Appendix~\ref{sec:unit-groups:casimir}, by an expression of the form
$\widetilde{\bf C}_2 = \sum_{jk} g^{jk} X_j X_k$.
This is all we needed in the derivation of equation~(\ref{omegaC2}) 
in the ${\rm SU}(2)$-case. The generalization to arbitrary $d$ 
hence reads
\begin{equation}
  \Label{eq:omega-casimir-2}
  \omega = \frac{1}{2} + \frac{\widetilde{C}_2(\pi_\mathbf{m}) -
  \widetilde{C}_2(\pi_\mathbf{n})}{2\widetilde{C}_2(\pi_N^+)}
\quad.
\end{equation}
The concrete form of $\widetilde{C}_2(\pi_\mathbf{m})$ as a function 
of the weights $\mathbf{m}$ is given in 
equation~(\ref{eq:casimir-3}), and will be needed only later.  
Since $\widetilde{C}_2(\pi_N^+)$ is a positive 
constant we have to maximize the function
\begin{equation}
  \Label{eq:F-def}
  W \ni (\mathbf{m},\mathbf{n}) \mapsto F(\mathbf{m},\mathbf{n}) =
  \widetilde{C}_2(\pi_\mathbf{m}) - \widetilde{C}_2(\pi_\mathbf{n})
  \in \Bbb{Z} 
\end{equation}
on its domain $W$. 

The first step in this direction is to reexpress
$F(\mathbf{m},\mathbf{n})$ in terms of the $\U(d)$ Casimir operators
${\bf C}_2$ and ${\bf C}_1^2$. Note in this context that although
equation (\ref{eq:omega-def-2}) is, as already stated, valid only for $X
\in \goth{su}(d)$ the representations $\pi_\mathbf{m}$ and
$\pi_\mathbf{n}$ are still $\U(d)$ representations Hence we can apply
the equation $\widetilde{\bf C}_2 = {\bf C}_2 - \frac{1}{d} {\bf
  C}_1^2$ given in Section \ref{sec:unit-groups:casimir}: 
\begin{equation}
\label{F0-functional}
  F(\mathbf{m},\mathbf{n}) = C_2(\pi_\mathbf{m}) - C_2(\pi_\mathbf{n})
  - \frac{1}{d}(C_1^2(\pi_\mathbf{m}) - C_1^2(\pi_\mathbf{n})).
\end{equation}
This rewriting is helpful, because the invariants $C_1$ turn out to 
be independent of the variational parameters:
Since $\pi_\mathbf{m} \subset \piB^\otm$, and 
$\partial\piB^\otm(\idty_d)=M\idty$, we also have 
$C_1(\pi_\mathbf{m})=M$. On the other hand, the existence of an 
intertwining isometry $V$ with $V \pi_N^+ = \pi_\mathbf{m} \otimes 
\pi_\mathbf{n}V$ implies
\begin{displaymath}
  V C_1(\pi_N^+) \Bbb{1} = V \partial \pi_N^+({\bf C}_1) =
  \left(\partial \pi_\mathbf{m}({\bf C}_1) \otimes \Bbb{1}_\mathbf{n}
  + \Bbb{1}_\mathbf{m} \otimes \partial \pi_\mathbf{n}({\bf
  C}_1)\right) V = \left(C_1(\pi_\mathbf{m}) \Bbb{1} +
  C_1(\pi_\mathbf{n}) \Bbb{1}\right) V 
\end{displaymath}
and therefore $C_1(\pi_N^+) = C_1(\pi_\mathbf{m}) +
C_1(\pi_\mathbf{n})$. Since $C_1(\pi_N^+) = N$ and
$C_1(\pi_\mathbf{m}) =M$ we get $C_1(\pi_\mathbf{n}) = N-M$. 
Inserting this into equation~(\ref{F0-functional}) we find the 
functional
\begin{equation}
  \Label{eq:F-F1}
 F(\mathbf{m},\mathbf{n}) = F_1(\mathbf{m},\mathbf{n}) - \frac{2MN -
 N^2}{d},  
\end{equation}
where only $F_1$ depends on the variational parameters, and is 
expressed explicitly (see equation \ref{eq:casimir-2}) as
\begin{equation}
  \Label{eq:f1-def}
  W \ni (\mathbf{m},\mathbf{n}) \mapsto F_1(\mathbf{m},\mathbf{n}) =
  C_2(\pi_\mathbf{m}) - C_2(\pi_\mathbf{n}) = \sum_{j=1}^d (m_j^2
  -n_j^2) + \sum_{k=1}^d (d-2k+1)(m_k - n_k) \in \Bbb{Z}
\end{equation}
which remains to be maximized over $W$.

To do this we have to express the constraints defining the domain 
$W$ more explicitly. We have already seen that $\mathbf{m} 
\in \Bbb{Z}^d_+$ has to satisfy the constraint $\sum_{j=1}^d m_j = 
M$. In addition we get, due to equation \ref{eq:tensprod-rep-1} $m_d > 
0$. To fix the constraints for $\mathbf{n}$ note that 
according to equation (\ref{eq:tensorprod-rep-4})
$\pi_N^+ \subset \pi_\mathbf{m} \otimes \pi_\mathbf{n}$ is  
equivalent to $\pi_\mathbf{m} \subset \pi_N^+ \otimes 
\pi_{\widetilde{\mathbf{n}}}$. Here we have introduced 
$\widetilde{\mathbf{n}} =(\widetilde{n}_1, \dots, \widetilde{n}_d) = 
(-n_d, \dots, -n_1)$ as a notation for the highest weight of the 
representation $\overline{\pi_\mathbf{n}}$ conjugate to
$\pi_\mathbf{n}$ (i.e.  $\overline{\pi_\mathbf{n}} =
\pi_{\widetilde{\mathbf{n}}}$). Now we can apply equation
(\ref{eq:tensorprod-rep-3}) to get  
\begin{displaymath}
  \pi_N^+ \subset \pi_\mathbf{m} \otimes \pi_\mathbf{n} \iff
  \widetilde{n}_k = m_k - \mu_k \mbox{ with } 0 \leq \mu_k \leq m_k -
  m_{k+1} \  \forall k=1,\dots,d-1 \mbox{ and } \sum_{k=1}^d \mu_k = 
N.
\end{displaymath}
In other words
\begin{displaymath}
  W = \{ (\mathbf{m},\mathbf{n}) \, | \, \widetilde{\mathbf{n}} =
  \mathbf{m} - \mathbf{\mu}, \mbox{ and } (\mathbf{m},\mathbf{\mu})
  \in W_1 \}
\end{displaymath}
with
\begin{displaymath}
  W_1 = \{ (\mathbf{m},\mathbf{\mu}) \in \Bbb{Z}^d_+ \times \Bbb{Z}^d
  \, | \, \sum_{k=1}^d m_k = M, \ \sum_{k=1}^d \mu_k = N \mbox{ and }
  0 \leq \mu_k \leq m_k - m_{k+1} \ \forall k = 1, \dots, d-1 \}.
\end{displaymath}
The function $F_1$ can now be re-expressed in terms the new variables
$(\mathbf{m},\mathbf{\mu})$. To this end note that
$C_2(\pi_\mathbf{n}) = C_2(\overline{\pi_\mathbf{n}}) =
C_2(\pi_{\widetilde{\mathbf{n}}})$. Hence we have
\begin{displaymath}
  F_1(\mathbf{m},\mathbf{n}) 
      = F_1(\mathbf{m},\widetilde{\mathbf{n}}) 
      = F_1(\mathbf{m}, \mathbf{m} - \mathbf{\mu})
\end{displaymath}
and therefore with equation (\ref{eq:f1-def}):
\begin{equation}
  \Label{eq:F1-F2}
  F_1(\mathbf{m},\widetilde{\mathbf{n}}) 
    = \sum_{k=1}^d \mu_k (2m_k -2k  -\mu_k) + (d+1) \sum_{k=1}^d \mu_k 
    = F_2(\mathbf{m},\mathbf{\mu}) +  (d+1)N 
\end{equation}
with the new function
\begin{equation}
  \Label{eq:f2-def}
  W_1 \ni (\mathbf{m},\mathbf{\mu}) \mapsto
  F_2(\mathbf{m},\mathbf{\mu}) = \sum_{k=1}^d \mu_k (2m_k -2k
  -\mu_k) \in \Bbb{Z}.
\end{equation}
Hence we have reduced our problem to the following Lemma:

\proclaim\XVarLemma\ Lemma. 
The function $F_2: W_1 \to \Bbb{Z}$ defined in equation
(\ref{eq:f2-def}) attains its maximum for and only for 
\begin{displaymath}
    \mathbf{m}_{\rm max}=(M,0,\ldots,0) 
\qquad\text{and}\qquad
    \mathbf{\mu}_{\rm max}
               =\cases{(N,0,\ldots,0)     & for $N\leq M$\cr
                       (M,0,\ldots,0,N-M) &  for $N\geq M$.}
\end{displaymath}

\noindent {\it Proof:}
We consider a number of cases in each of which we apply a different 
strategy for increasing $F_2$. In these procedures we consider $d$ to 
be a variable parameter, too, because if $\mu_d=m_d=0$, the further 
optimization will be treated as a special case of the same problem 
with $d$ reduced by one. 

{\it Case A:} $\mu_d>0$, $\mu_i<m_i-m_{i+1}$ for some $i<d$.\\
In this case we apply the substitution $\mu_i\mapsto(\mu_i+1)$, 
$\mu_d\mapsto(\mu_d-1)$, which leads to the change 
\begin{displaymath}
    \delta F_2= 2\bigl(-\mu_i+\mu_d +(d-i-1)+ m_{i+1}-m_d\bigr)
       \geq 2\bigl(\mu_d +(d-i-1)\bigr)>0
\end{displaymath}
in the target functional. In this way we proceed until either all 
$\mu_i$ with $i<d$ satisfy the upper bound with equality (Case~B 
below) or $\mu_d=0$, i.e., Case~C or Case~D applies. 

{\it Case B:} $\mu_d>0$, $\mu_i=m_i-m_{i+1}$ for all
  $i<d$. 
In this case all $\mu_k$, including $\mu_d$ are determined by the 
$m_k$ and by the normalization ($\mu_d=N-m_1+m_d$). Inserting these 
values into $F_2$, and using the normalization conditions, we get
$F_2(\mathbf{m},\mathbf{n}) = F_3(\mathbf{m}) -2(M+dN)-N^2$ with
\begin{eqnarray*}
    F_3({\bf m})
      &=&2(N+d) m_1 \\
\text{constrained by} \qquad
    m_1\geq\cdots\geq m_d\geq0&,& \sum_km_k=M,  \qquad
\text{and }             m_1-m_d\leq N.
\end{eqnarray*}
This defines a variational problem in its own right. Any step 
increasing $m_1$ at the expense of some other $m_k$ increases $F_2$. 
This process terminates either, when $M=m_1$, and all other $m_k=0$. 
This is surely the case for $M<N$, because then 
$\mu_d=N-m_1+m_d\geq N-M>0$. This is already the final result 
claimed in the Lemma. On the other hand, the process may terminate 
because $\mu_d$ reaches $0$ or would become negative. 
In the former case we get $\mu_d=0$, and hence Case~C or Case~D. 
The latter case (termination at $\mu_d=1$) may occur because the 
transformation $m_1\mapsto(m_1+1)$, $m_d\mapsto(m_d-1)$ changes 
$\mu_d=N-m_1+m_d$ by $-2$. There are two basic situations in which 
changing both $m_1$ and $m_d$ is the only option for maximizing 
$F_3$, namely $d=2$ and $m_1=m_2=\cdots=m_d$. The first case is 
treated below as Case~E. In the latter case we have $1=N-m_1+m_d=N$. 
Then the overall variational problem in the Lemma is trivial, 
because only one term remains, and one only has to maximize the
quantity $2m_k-2k-1$, with trivial maximum at $k=1$, $m_1=M$. 

{\it Case C:} $\mu_d=0$, $m_d>0$. 
For $\mu_d=0$, the number $m_d$ does not enter in the function
$F_2$. Therefore, the move $m_d\mapsto0$ and $m_1\mapsto m_1+m_d$, 
increases $F_2$ by $\mu_1m_d\geq0$. Note that this is always 
compatible with the constraints, and we end up in Case~D. 

{\it Case D:} $\mu_d=0$, $m_d=0$, $d>2$. 
Set $d\mapsto(d-1)$. Note that we could now use the extra constraint 
$\mu_{d'}\leq m_{d'}$, where $d'=d-1$. We will {\it not} use it, so 
in principle we might get a larger maximum. However, since we do 
find a maximizer satisfying all constraints, we still get a valid 
maximum. 

{\it Case E:} $d=2, \mu_1=m_1-m_2, \mu_2=1$.
In this case $\mathbf{m} = (m_1,m_2)$ is completely fixed by the
constraints. We have: $m_1+m_2=M$ and $\mu_1 + \mu_2 = m_1-m_2+1 = N$
hence $m_1-m_2=N-1$. This implies $2m_1 = M+N-1$, $2m_2 = M - N +1$
and since $m_2 \geq 0$ we get $M\geq N-1$. If $M=N-1$ holds we get
$m_1 = N-1 = M$, $ m_2 = 0$ and consequently $\mu_1 = N-1$. Together
with $\mu_2 = 1 = N-M$ these are exactly the parameters where $F_2$
should take its maximum according to the Lemma. Hence assume $M \geq
N$. In this case $\mu_2=1$ implies that $F_2$ becomes $NM-3N-4$, which
is, due to $M\geq N$, strictly smaller than $F_2(M,0;N,0) =
2MN-N^2-2N$. 

{\it Uniqueness:} In all cases just discussed the manipulations
described lead to a strict increase of $F_2(\mathbf{m},\mu)$ as long as
$(\mathbf{m},\mu) \not= (\mathbf{m}_{\rm max}, \mu_{\rm max})$
holds. The only exception is Case~C with $\mu_1=0$. In this situation
there is a $1 < k < d$ with $\mu_k>0$. Hence we can apply the maps $d
\mapsto d-1$ (Case~D) and $m_d\mapsto0$ and $m_1\mapsto m_1+m_d$
(Case~C) until we get $\mu_d \not= 0$ (i.e. $d$ reaches $k$). Since
$\mu_1=0$ the corresponding $(\mathbf{m},\mu)$ is not equal to
$(\mathbf{m}_{\rm max}, \mu_{\rm max})$. Therefore we can apply one
of manipulations described in Case~A, Case~B or Case~E which leads to
a strict increase of $F_2(\mathbf{m},\mu)$. This shows that
$F_2(\mathbf{m},\mu) < F_2(\mathbf{m}_{\rm max},\mu_{\rm max})$ as
long as $(\mathbf{m}, \mu) \not= (\mathbf{m}_{\rm max},\mu_{\rm max})$
holds. Consequently the maximum is unique. \QED

With this result and the equations (\ref{eq:omega-casimir-2}),
(\ref{eq:F-def}), (\ref{eq:F-F1}), (\ref{eq:F1-F2}) and
(\ref{eq:f2-def}) we can easily calculate $\omega_{\rm max}$:
\begin{displaymath}
  \omega_{\rm max} = \omega(\widehat{T}) = \frac{M+d}{N+d}
\end{displaymath}
and with (\ref{Delomega}) we get $\Delta(T) \geq \Delta(\widehat{T})$
with $\Delta(\widehat{T})$ from Theorem \XTheorem.

\subsection{Proving uniqueness}
\label{sec:unique}

One part of the uniqueness proof is already given above: there is 
only one optimal {\em covariant} cloning map, namely
$\widehat{T}$. This follows easily from the uniqueness of the maximum
found in Lemma \XVarLemma\ and from the fact that the representation
$\pi^+_N$ is contained exactly once in the tensor product $\pi^+_M
\otimes \overline{\pi^+_{M-n}}$ (see equation
\ref{eq:tensorprod-rep-3} and the discussion in Subsection
\ref{sec:covex-decompo}). 

Suppose now that $T$ is a non-covariant cloning map, which also 
attains the best value: 
$\Delta_{\rm one}(T)=\Delta_{\rm one}(\widehat{T})$. Then we may 
consider the average of $\overline{T}$ of $T$ (see 
equation~(\ref{mean})), which is also 
optimal and, in addition, covariant. Therefore 
$\overline{T}=\widehat{T}$. The uniqueness part of the proof thus
follows immediately from the following proposition:

\proclaim\XUniqueProp\ Proposition.
Each completely positive, unital map $T: {\cal B}({\cal H}^\otm)
\to {\cal B}({\cal H}_+^\otn)$ satisfying the equation $\overline{T} =
\widehat{T}$ equals $\widehat{T}$.

\noindent {\it Proof:} 
We trace back this statement to the main theorem of \cite{part1}. To
this end note that $\overline{T}=\widehat{T}$ implies the equivalent
equation for the preduals: 
\begin{displaymath}
  \overline{T_*} = \int \tau_u T_* du = \widehat{T}_*
\end{displaymath}
where $\tau_u$ acts on $T_*$ by:
\begin{displaymath}
  \tau_u T_*(\sigma) = \piB^\otm(u)^* T_*\left(\pi^+_N(u) \sigma
  \pi^+_N(u)^*\right)\piB^\otm(u).
\end{displaymath}
Furthermore we know from the main theorem of \cite{part1} that
$\tr\left(\sigma^\otm T_*(\sigma^\otn)\right) \leq \frac{d[N]}{d[M]}$
is true for all pure states $\sigma \in {\cal B}_*({\cal H})$ and that
equality holds iff $T=\widehat{T}$. Consequently we have
\begin{displaymath}
  \int \left( \frac{d[N]}{d[M]} - \tr\left(\sigma^\otm \tau_u
  T(\sigma^\otn) \right) \right) du = \frac{d[N]}{d[M]} - \tr \left(
  \sigma^\otm \overline{T_*}(\sigma^\otn)\right) = \frac{d[N]}{d[M]} -
  \tr \left( \sigma^\otm \widehat{T}_*(\sigma^\otn)\right) = 0.
\end{displaymath}
Since the integral on the left hand site of this equation is taken
over positive quantities the integrand has to vanish for all values of
$u \in \U(d)$. This implies $\tr\left(\sigma^\otm T(\sigma^\otn)
\right) = \frac{d[N]}{d[M]}$ for all pure states $\sigma \in {\cal
  B}_*({\cal H})$. However this is, according to \cite{part1}, only
possible if $T = \widehat{T}$.
\QED 

\section*{Acknowledgements}

This paper is a response to the many discussions about the cloning 
problem one of us (R.F.W.) had with participants of the ISI Workshop
on Quantum Computing in Torino in July 1997. We would like to thank
D. Sondermann for a critical reading of the manuscript. 

\appendix

\section{Representations of unitary groups}
\label{sec:unit-groups}

Throughout this paper many arguments from representation
theory of unitary groups are used. In order to fix the notation and to
state the most relevant theorems we will recall in this appendix some
well known facts from representation theory of Lie groups. General
references are the books of Barut and Raczka \cite{BARA}, Zhelobenko
\cite{ZHELOBENKO} and Simon \cite{Simon}.

\subsection{The groups and their Lie algebras}
\label{sec:unit-groups:lie-algs}

Let us consider first the group $\U(d)$ of all complex $d \times d$
unitary matrices. Its Lie algebra $\goth{u}(d)$ can be identified with
the Lie algebra of all anti-hermitian $d \times d$ matrices. The
exponential function is then given by the usual matrix exponential $X
\mapsto \exp(iX)$. $\goth{u}(d)$ is a real Lie algebra. Hence
we can consider its complexification $\goth{u}(d) \otimes \Bbb{C}$
which coincides with the set of all $d\times d$ matrices and at the
same time with the Lie algebra $\goth{gl}(d,\Bbb{C})$ of the general
linear group $\GL(d,\Bbb{C})$. In other words $\goth{u}(d)$ is a real
form of $\goth{gl}(d, \Bbb{C})$. A basis of $\goth{gl}(d,\Bbb{C})$
is given by the matrices $E_{jk} = |j\rangle \langle
k|$.

The set of elements of $\U(d)$ with determinant one forms the subgroup
$\SU d$ of $\U(d)$. Its Lie algebra $\goth{su}(d)$ is the subalgebra
of $\goth{u}(d)$ consisting of the elements with zero trace. Hence
the complexification $\goth{su}(d) \otimes \Bbb{C}$ of $\goth{su}(d)$
is the Lie algebra of trace-free matrices and coincides therefore with
the Lie algebra $\goth{sl}(d, \Bbb{C})$ of the special linear group
$\SL(d, \Bbb{C})$. As well as in the $\U(d)$ case this means that
$\goth{su}(d)$ is a real from of $\goth{sl}(d, \Bbb{C})$. The matrices
$E_{jk}$ are no longer a basis for $\goth{sl}(d,\Bbb{C})$ since the
$E_{jj}$ are not trace free. Instead we have to consider $E_{jk}$, $j
\not= k$ and $H_j = E_{jj} - E_{j+1,j+1}$, $j=1, \dots, d-1$. The
difference between $\goth{sl}(d, \Bbb{C})$ and $\goth{gl}(d, \Bbb{C})$
is exactly the center of $\goth{gl}(d, \Bbb{C})$, i.e. all complex
multiples of the identity matrix. In other words we have $\goth{gl}(d,
\Bbb{C}) = \goth{sl}(d, \Bbb{C}) \oplus \Bbb{C} \Bbb{1}$.  A similar
result holds for the real forms: $\goth{u}(d) = \goth{su}(d) \oplus
\Bbb{R} \Bbb{1}$. 

The (real) span of all $iE_{jj}$, $j=1, \dots, d$ is a subalgebra of
$\goth{u}(d)$ which is maximal abelian, i.e. a Cartan subalgebra of
$\goth{u}(d)$. We will denote it in the following by $\goth{t}(d)$ and
its complexification by $\goth{t}_{\Bbb{C}}(d) \subset
\goth{gl}(d,\Bbb{C})$. The intersection of $\goth{t}(d)$ with
$\goth{su}(d)$ results in a Cartan subalgebra $\goth{st}(d)$ of
$\goth{su}(d)$. We will denote the complexification by
$\goth{st}_{\Bbb{C}}(d)$. Again the two algebras $\goth{t}(d)$ and
$\goth{st}(d)$ differ by the center of $\goth{u}(d)$ i.e. $\goth{t}(d)
= \goth{st}(d) \oplus \Bbb{R} \Bbb{1}$ and $\goth{t}_{\Bbb{C}}(d) = 
\goth{st}_{\Bbb{C}}(d) \oplus \Bbb{C} \Bbb{1}$ in the complexified case.

\subsection{Representations}
\label{sec:unit-groups:reps}

Consider now a finite-dimensional\footnote{All representations in this
  paper are finite dimensional.} representation $\pi: \U(d) \to
\GL(N,\Bbb{C})$ of $\U(d)$. It is characterized uniquely by the
corresponding representation $\partial \pi: \goth{u}(d) \to
\goth{gl}(N,\Bbb{C})$ of its Lie algebra, i.e. we have $\pi(\exp(X))
= \exp(\partial \pi(X))$. The representation $\partial \pi$ can
be extended by complex linearity to a representation of $\goth{gl}(d,
\Bbb{C})$ which we will denote by $\partial \pi$ as well. Hence
$\partial \pi$ leads to a representation $\pi$ of the group
$\GL(d,\Bbb{C})$. Similar notations we will adopt for representations
of $\SU d$ and $\SL(d, \Bbb{C})$.    

Assume now that $\pi$ is an irreducible representation of
$\GL(d, \Bbb{C})$. An infinitesimal weight of $\pi$ (or simply a
weight in the following) is an element $\lambda$ of the dual of
$\goth{t}_{\Bbb{C}}^*(d)$ of $\goth{t}_{\Bbb{C}}(d)$ such that
$\partial \pi(X) x = \lambda(X) x$ holds for all $X \in
\goth{t}_{\Bbb{C}}(d)$ and for a nonvanishing $x \in \Bbb{C}^N$. The
linear subspace $V_\lambda \subset \Bbb{C}^N$ of all such $x$ is
called the weight subspace of the weight $\lambda$. The set of weights
of $\pi$ is not empty and, due to irreducibility, there is exactly one
weight ${\bf m}$, called the highest weight, such that $\partial
\pi(E_{jk}) x = 0$ for all $x$ in the weight subspace of ${\bf m}$ and
for all $j,k = 1, \dots, d$ with $j < k$. The representation $\pi$ is
(up to unitary equivalence) uniquely determined by its highest
weight. On the other hand the weight ${\bf m}$ is uniquely determined
by its values ${\bf m}(E_{jj}) = m_j$ on the basis $E_{jj}$ of
$\goth{t}_{\Bbb{C}}(d)$. We will express this fact in the following as
``${\bf m} = (m_1, \dots, m_d)$ is the highest weight of the
representation $\pi$''. For each analytic representation of
$\GL(d,\Bbb{C})$ the $m_j$ are integers satisfying the inequalities
$m_1 \geq m_2 \geq \dots \geq m_d$ and the converse is also true: each
family of integers with this property defines the highest weight of an
analytic, irreducible representation of $\GL(d, \Bbb{C})$.

In a similar way we can define weights and highest weights for
representations of the group $\SL(d, \Bbb{C})$ as linear forms on the
Cartan subalgebra $\goth{st}_{\Bbb{C}}(d)$. As in the
$\GL(d,\Bbb{C})$-case an irreducible representation $\pi$ of $\SL(d,
\Bbb{C})$ is characterized uniquely by its highest weight ${\bf
  m}$. However we can not evaluate ${\bf m}$ on the basis $E_{jj}$
since these matrices are not trace free. One possibility is to
consider an arbitrary extension of ${\bf m}$ to the algebra
$\goth{t}_{\Bbb{C}}(d) = \goth{st}_{\Bbb{C}}(d) \oplus \Bbb{C}
\Bbb{1}$. Obviously this extension is not unique. Therefore the values
${\bf m}(E_{jj}) = m_j$ are unique only up to an additive
constant. To circumvent this problem we will use usually the
normalization condition $m_d = 0$. In this case the integer $m_j$
corresponds to the number of boxes in the $j^{th}$ row of the Young
tableau usually used to characterize the irreducible representation
$\pi$. Another possibility to describe the weight ${\bf m}$ is to use
the basis $H_j$ of $\goth{st}_{\Bbb{C}}(d)$. We get a sequence of
integers $l_j = {\bf m}(H_j)$, $j = 1, \dots, d-1$. They are related
to the $m_j$ by $l_j = m_j-m_{j+1}$. Each sequence $l_1, \dots,
l_{d-1}$ defines the highest weight of an irreducible representation
of $\SL(d, \Bbb{C})$ iff the $l_j$ are positive integers.   

Finally consider the representation $\Bar \pi$ conjugate to $\pi$,
i.e. $\overline{\pi}(u) =\overline{\pi(u)}$. If $\pi$
is irreducible the same is true for $\Bar \pi$. Hence $\Bar \pi$
admits a highest weight which is given by $(-m_d, -m_{d-1}, \dots,
-m_1)$. If $\pi$ is a $\SU d$ representation we can apply the
normalization $m_d=0$. Doing this as well for the conjugate
representation we get $(m_1, m_1 - m_{d-1}, \dots, m_1 - m_2, 0)$. In
terms of Young tableaus this corresponds to the usual rule to
construct the tableau of the conjugate representation: Complete the
Young tableau of $\pi$ to form a $d \times m_1$ rectangle. The
complementary tableau rotated by $180^\circ$ is the Young tableau of
$\Bar \pi$. 

\subsection{Tensor products of representations}
\label{sec:unit-groups:tensor-prod}

Consider now two finite dimensional irreducible representations
$\pi_{\bf m}, \pi_{\bf n}$ of $\U(d)$ with highest weights ${\bf m},
{\bf n}$. Their tensor product $\pi_{\bf m} \otimes \pi_{\bf n}$ is
completely reducible. If $r_\pi$ denotes the multiplicity of the
irreducible representation $\pi$ in $\pi_{\bf m} \otimes \pi_{\bf n}$
then this means that $\pi_{\bf m} \otimes \pi_{\bf n} = \bigoplus_\pi
r_\pi \pi$. Hence to decompose the representation $\pi_{\bf m} \otimes
\pi_{\bf n}$ we have to compute the  integer valued functions $({\bf
  m}, {\bf n}) \mapsto r_\pi({\bf m},{\bf n})$. There are several
general schemes to do this (see e.g. \cite[Ch. XII]{ZHELOBENKO}).
However we are only interested in the following special cases.   
The highest weight of the representation $\pi_{\bf 1}: \U(d) \ni U \mapsto U
\in \GL(d, \Bbb{C})$ (denoted $\piB$ in Section \ref{sec:reduction}) is
${\bf 1} = (1, 0, \dots, 0)$. Consider the $N-$fold tensor product of
this representation It can be decomposed as follows 
\begin{equation}
  \label{eq:tensprod-rep-1}
  \pi_{\bf 1}^\otn = \sum_{m_1+\dots+m_d=N \atop m_d \geq 0}
  r(m_1,\dots,m_d) \pi_{m_1,\dots,m_d} 
\end{equation}
where $\pi_{m_1,\dots,m_d}$ denotes the irreducible representation
with highest weight $(m_1,\dots,m_d)$. The coefficients $r(m_1, \dots,
m_d)$ are determined by the following recurrence relation: 
\begin{equation}
  \label{eq:tensorprod-rep-2}
  r(m_1,\dots,m_d) = r(m_1-1,\dots,m_d) + r(m_1,m_2-1,\dots,m_d) +
  \dots + r(m_1,\dots,m_d-1).
\end{equation}

Consider now the $N-$fold symmetric tensor product of
$\pi_{\bf 1}$ (denoted $\pi^+_N$ in Section \ref{sec:reduction}). It is
irreducible with highest weight $N {\bf 1} = (N, 0, \dots, 0)$ (hence
$\pi_N^+ = \pi_{N {\bf 1}}$). The tensor product of this
representation with an arbitrary irreducible representation $\pi_{\bf
  m}$ (with highest weight ${\bf m} = (m_1,\dots,m_d)$) is
\begin{equation}
  \label{eq:tensorprod-rep-3}
  \pi_{N{\bf 1}} \otimes \pi_{\bf m} = \sum_{0 \leq \mu_{k+1} \leq
  m_k - m_{k+1} \atop \mu_1 + \dots + \mu_d = N} \pi_{m_1 
  + \mu_1, \dots, m_d + \mu_d}.
\end{equation}
From this equation we also get a condition for $\pi_{N{\bf 1}}$ to be
contained in an arbitrary tensor product $\pi_{\bf m} \otimes \pi_{\bf
  n}$ which we need in Section \ref{sec:maxOmega-general}: For arbitrary
weights ${\bf m}, {\bf n}, {\bf p}$ we have
\begin{equation}
  \label{eq:tensorprod-rep-4}
  \pi_{\bf m} \subset \pi_{\bf n} \otimes \pi_{\bf p} \iff
  \pi_{\bf n} \subset \overline{\pi_{\bf p}} \otimes \pi_{\bf m}
\end{equation}

If two irreducible representations $\pi_{\bf m}, \pi_{\bf n}$ of $\SU
d$ are given we can characterize them, as described above, by their
highest weights ${\bf m} = (m_1, \dots, m_d)$ and ${\bf n}=(n_1,
\dots, n_d)$ using the normalizations $m_d = 0$ and $n_d=0$. After
applying the stated theorems to the tensor product of the
corresponding $\U(d)$ representations we can restrict the summands in
the resulting spectral decomposition back to $\SU d$, i.e. we
renormalize the heighest weigths $(m_1, \dots, m_d)$ to the $m_d = 0$
case. 

\subsection{Nondegeneracy of $\goth{su}(d)$}
\label{sec:nondeg-pf}

We are now ready to discuss the group theoretic part of the proof 
of our main theorem, i.e. Lemma \XLnondeg\ which we have only stated
in Section \ref{sec:finding}. According to Def. \XDnondeg\ we have to
show that each linear operator $\Lambda : \goth{su}(d) \to {\cal
  H}_+^\otn$ with the covariance property 
\begin{equation}
  \label{eq:covar-prop}
  \pi^+_N(g)\Lambda(X)\pi^+_N(g^{-1}) =\Lambda(gXg^{-1})  
\end{equation}
is of the form $\Lambda(X) = \lambda \partial\pi_+^N(X)$ with a
constant factor $\lambda$. Here $\pi^+_N$ is the irreducible
representation of $\SU d$ introduced in Section
\ref{sec:reduction}. (Hence we have $\pi^+_N = \pi_{N {\bf 1}}$ using
the notation introduced in Subsection
\ref{sec:unit-groups:tensor-prod} of this appendix.) 

To reformulate this statement note first that the map $g \mapsto
\pi_N^+(g) \, \cdot \, \pi_N^+(g^{-1})$ can be interpreted as a
unitary representation of $\SU d$ on the representation space ${\cal
  H}^\otn_+ \otimes {\cal H}^\otn_+$. In fact it is (unitarily
equivalent to) the tensor product $\pi_N^+ \otimes
\overline{\pi_N^+}$. Since $\SU d \ni g \mapsto g \,\cdot\, g^{-1} \in
{\cal B}(\goth{su}(d))$ is the adjoint representation of $\SU d$ this
implies that each map $X$ satisfying (\ref{eq:covar-prop}) intertwines
$\pi_N^+ \otimes \overline{\pi_N^+}$ and the adjoint representation
$\Ad$. Note second that the representation $\partial\pi^+_N$ of the
Lie algebra $\goth{su}(d)$ satisfies equation (\ref{eq:covar-prop}) in
an obvious way (with $\lambda =1$) hence we have to show that all such
intertwiners are proportional, or in other words that $\Ad$ is
contained in $\pi_N^+ \otimes \overline{\pi_N^+}$ exactly once.

Let us discuss now the tensor product $\pi_N^+ \otimes
\overline{\pi_N^+}$. The irreducible representation $\pi^+_N$ has
highest weight $(N,0,\dots,0)$ (see Section
\ref{sec:unit-groups:reps}) and consequently the highest weight of its
conjugate is $(N,\dots,N,0)$. We can apply now equation
(\ref{eq:tensorprod-rep-3}) which shows that the adjoint
representation whose highest weight is $(2,1,\dots,1,0)$ is contained in
$\pi_N^+ \otimes \overline{\pi_N^+}$ exactly ones. This shows together
with our previous discussion that $\goth{su}(d)$ is nondegenerate in ${\cal
  H}^\otn_+$ with respect to $\pi^+_N$.

\subsection{The Casimir invariants}
\label{sec:unit-groups:casimir}

To each Lie algebra $\goth{g}$ we can associate its universal
enveloping algebra $\goth{G}$. It is defined as the quotient of the
full tensor algebra $\bigoplus_{n \in \Bbb{N}_0} \goth{g}^\otn$ with
the two sided ideal $\goth{I}$ generated by $X \otimes Y - Y \otimes X
- [X,Y]$, i.e. $\goth{G}$ is an associative algebra. The original Lie
algebra $\goth{g}$ can be embedded in its envelopping algebra
$\goth{G}$ by $\goth{g} \ni X \mapsto X + \goth{I} \in \goth{G}$. The
Lie bracket is then simply given by $[X,Y] = XY - YX$. Moreover
$\goth{G}$ is algebraically generated by $\goth{g}$ and $\Bbb{1}$.
Hence each representation $\partial \pi$ of $\goth{g}$ generates a
unique representation $\partial \pi$ of $\goth{G}$ simply by $\partial
\pi(X_1 \cdots X_k) = \partial \pi(X_1) \cdots \partial \pi(X_k)$. If
$\partial \pi$ is irreducible the induced representation $\partial
\pi$ is irreducible as well. 

We are interested not in the whole algebra but only in its center
$\goth{Z}(\goth{G})$, i.e. the subalgebra consisting of all $Z \in
\goth{G}$ commuting with all elements of $\goth{G}$. The elements of
$\goth{Z}(\goth{G})$ are called central elements or Casimir elements.
If $\partial \pi$ is a representation of $\goth{G}$ the representatives
$\partial \pi(Z)$ of Casimir elements commute with all other
representatives $\partial \pi(X)$. This implies for irreducible
representations that all $\partial \pi(Z)$ are multiples of the identity. 

Consider now the case $\goth{g} = \goth{gl}(d, \Bbb{C})$. In this case
we can identify the envelopping algebra $\goth{G}$ with the set of all
left invariant differential operators on $\GL(d,\Bbb{C})$ (a similar
statement is true for any Lie group). Of special interest for us are
the Casimir elements belonging to operators of first and second
order. Using the standard basis $E_{ij}$ of $\goth{gl}(d, \Bbb{C})$
introduced in Section \ref{sec:unit-groups:lie-algs} they are given by
\begin{displaymath}
{\bf C}_1 = \sum_{j=1}^d E_{jj} \mbox{ and } {\bf C}_2 =
\sum_{j,k=1}^d E_{jk}E_{kj}.   
\end{displaymath}
 Of course ${\bf C}_1^2$ is as well of second order and
it is linearly independent of ${\bf C}_2$. Hence each second order
Casimir element of $\goth{G}$ is a linear combination of ${\bf C}_2$
and ${\bf C}_1^2$. 

If $\partial \pi$ is an irreducible representation of
$\goth{gl}(d,\Bbb{C})$ with highest weight $(m_1, \dots, m_d)$ it
induces, as described above, an irreducible representation $\partial
\pi$ of $\goth{G}$ and the images of $\partial \pi({\bf C}_1)$ and
$\partial \pi({\bf C}_2)$ are multiples of the 
identity, i.e. $\partial \pi({\bf C}_1) = C_1(\pi) \Bbb{1}$ and
$\partial \pi({\bf C}_2) = C_2(\pi)\Bbb{1}$ with 
\begin{equation}
  \label{eq:casimir-1}
  C_1(\pi) = \sum_{j=1}^d m_j \mbox{ and } C_2(\pi)=\sum_{j=1}^d m_j^2
  + \sum_{j<k} (m_j - m_k).
\end{equation}

Let us discuss now the Casimir elements of $\SL(d,\Bbb{C})$. Since
$\SL(d,\Bbb{C})$ is a subgroup of $\GL(d,\Bbb{C})$ its enveloping
algebra $\goth{S}$ is a subalgebra of
$\goth{G}$. However the corresponding Lie algebras differ 
only by the center of $\goth{gl}(d, \Bbb{C})$. Hence the center
$\goth{Z}(\goth{S})$ of $\goth{S}$ is a subalgebra of
$\goth{Z}(\goth{G})$. Since $\goth{sl}(d,\Bbb{C})$ is simple there is
no first order Casimir element and there is only one second order
Casimir element $\widetilde{\bf C}_2$ which is therefore a linear
combination $\widetilde{\bf C}_2 = {\bf C}_2 + \alpha {\bf C}_1^2$ of
${\bf C}_1^2$ and ${\bf C}_2$. Obviously the factor $\alpha$ is
uniquely determined by the condition that the expression
\begin{equation}
  \label{eq:casimir-2}
  \widetilde{C}_2(\pi) = C_1(\pi) + \alpha C_1^2(\pi) = \sum_{j=1}^d
  m_j^2 + \sum_{j<k} (m_j - m_k) + \alpha \left(\sum_{j=1}^d
  m_j\right)^2 
\end{equation}
with $\partial \pi(\widetilde{\rm C}_2) = \widetilde{C}_2(\pi) \Bbb{1}$ is
invariant under the renormalization $(m_1, \dots, m_d) \mapsto (m_1 +
\mu, \dots, m_d + \mu)$ with an arbitrary constant
$\mu$. Straightforward calculations show that $\alpha =
-\frac{1}{d}$. Hence we get $\widetilde{C}_2 = C_2 - \frac{1}{d} C_1^2$
and 
\begin{equation}
  \label{eq:casimir-3}
   \widetilde{C}_2(\pi) = \frac{1}{d}\left((d-1) \sum_{j=1}^d m_j^2 - 
  \sum_{j\not=k}^d m_j m_k + d \sum_{j<k} (m_j - m_k) \right).
\end{equation}
Alternatively $\widetilde{\bf C}_2$ can be expressed in terms of a
basis $(X_j)_j$ of $\goth{sl}(d,\Bbb{C})$. In fact there is a
symmetric second rank tensor $g^{jk}X_j \otimes X_k \in
\goth{sl}(d,\Bbb{C}) \otimes \goth{sl}(d,\Bbb{C})$ such that
$\widetilde{\bf C}_2$ coincides with the equivalence class of
$g^{jk}$ in $\goth{S}$. In other words $\widetilde{\bf
  C}_2 = \sum_{jk} g^{jk} X_j X_k$ holds which leads to 
\begin{equation}
  \label{eq:casimir-4}
  \widetilde{C}_2(\pi) \Bbb{1} = \sum_{jk} g^{jk} \partial \pi(X_i)
  \partial \pi(X_j) 
\end{equation}
for an irreducible representation $\pi$ of $\SU d$.

\section{Stinespring theorem for covariant cp-maps}
\label{sec:covar-stspr}

In this appendix we will state the covariant version of Stinespring's
theorem \cite{Scutaru} which we have used in the proof of Theorem
\XTheorem. However, as in the rest of the paper, we will restrict the
discussion to finite dimensional Hilbert spaces (i.e. only cp-maps
between finite von Neumann factors are considered). 

\proclaim \XStSprThm\ Theorem. Let $G$ be a group with finite
dimensional unitary representations $\pi_i: G \to {\cal B}({\cal H}_i)$ 
($i=1,2$), and $T:{\cal B}({\cal H}_2) \to {\cal B}({\cal H}_1)$ 
a completely positive map
with the covariance property $\pi_1(g)T(X)\pi_1(g)^* =
T(\pi_2(g)X\pi_2(g)^*)$.
\begin{enumerate}
\item 
  Then there is another finite dimensional unitary representation
  $\tilde\pi: G \to {\cal B}(\tilde{\cal H})$ and an intertwiner $V:
  {\cal H}_1 \to {\cal H}_2 \otimes \tilde{\cal H}$ with $V \pi_1(g) =
  \pi_2 \otimes \tilde \pi V$ such that $T(X) = V^*(X\otimes\Bbb{1})V$
  holds. 
\item 
  If $T = \sum_\alpha T^\alpha$ is a decomposition of $T$ in completely
  positive terms, there is a decomposition 
  $\idty=\sum_\alpha F^\alpha$ of the identity operator on 
  $\tilde{\cal H}$ into positive operators 
  $F^\alpha \in {\cal B}(\tilde{\cal H})$ with
  $[F^\alpha,\tilde\pi(g)]=0$ such that $T^\alpha(X) = V^* (X \otimes
  F^\alpha)V$
\end{enumerate}

We only sketch the main ideas of the proof.  
The first step is Stinespring's theorem in its general form 
\cite{Stinespring}: 
There exists a representation $\eta : {\cal B}({\cal H}_2) \to {\cal 
B}({\cal K})$ of the C*-algebra ${\cal B}({\cal H}_2)$ on a Hilbert 
space ${\cal K}$ and a bounded operator $V: {\cal H}_1 \to {\cal K}$ 
such that $T(X) = V^*\eta(X)V$ holds. Up to unitary equivalence there 
is exactly one such triple $({\cal K}, V, \pi)$ such that the 
vectors $\pi(A)V\psi \in {\cal K}$ with $\psi \in {\cal H}_1$ and 
$A\in{\cal B}({\cal H}_2)$ span ${\cal K}$.

It is this uniqueness, from which the representation $\tilde\pi$ of 
$G$ is constructed. Indeed, the objects  $V_g=V\pi_1(g)$, and 
$\eta_g(X)=\eta(\pi_2(g)X\pi_2(g)^*)$ form a Stinespring dilation of 
the completely positive map 
$T_g(X)=\pi_1(g)^*T(\pi_2(g)X\pi_2(g)^*)\pi_1(g)$, which by 
covariance is equal to $T$. Hence by ``uniqueness up to unitary 
equivalence'' there is a unique unitary operator $U_g\in\B(\K)$ such 
that $V_g=V\pi_1(g)=U_gV$, and 
$\eta_g(X)=\eta(\pi_2(g)X\pi_2(g)^*)=U_g\eta(X)U_g^*$. This can be 
simplified a bit further by the observation that according to the 
second equation the operators $\tilde U_g=\eta(\pi_2(g))^*U_g$ 
commute with all $\eta(X)$. It is easy to see that the $U_g$ are a 
representation, and hence so is $\tilde U$: we have
$\tilde U_{g}\tilde U_{h}
   =\eta(\pi_2(g))^*U_g\eta(\pi_2(h)^*)U_h
   =\eta(\pi_2(g))^*\eta(\pi_2(g)\pi_2(h)^*\pi_2(g)^*)U_gU_h
   =\eta(\pi_2(g)^*\pi_2(g)\pi_2(h)^*\pi_2(g)^*)U_{gh}
   =\eta(\pi_2(gh)^*)U_{gh}=\tilde U_{gh}$. 

For a proof of part (1) we now only need to invoke the observation 
that all representations of
${\cal B}({\cal H}_2)$ are of the form $\eta \simeq {\rm id} \otimes
\Bbb{1}$ with ${\cal K} = {\cal H}_2 \otimes \tilde {\cal H}$. (Here 
``$\simeq$'' denotes a unitary equivalence, which we will include as 
a factor in $V$. Since $\tilde U_g$ commutes with all 
$\eta(X)=X\otimes\idty$, it is of the form 
$\tilde U_g=\idty\otimes\tilde\pi(g)$, which proves the assertion. 

The second part of the theorem stated for a trivial group $G=\lbrace 
e\rbrace$ is also known as  the Radon-Nikodyn theorem coming with 
the Stinespring theorem. In general it asserts the existence of a 
partition of the identity operator on ${\cal K}$ into operators 
$\tilde F^\alpha$ commuting with all $\eta(X)$, giving the 
decomposition of $T$ as $T^\alpha=V^*\eta(X)F^\alpha V$. 
Again, we can write these as $\tilde F^\alpha=\idty\otimes 
F^\alpha$. Since the $F^\alpha V$ are uniquely determined by the 
$T^\alpha$, it is easy to see that covariance of $T^\alpha$ is 
equivalent to $F^\alpha=\tilde\pi_gF^\alpha\tilde\pi_g^*$.

\end{document}